\documentclass[conference]{IEEEtran}
\usepackage{cite}
\usepackage{amsmath,amssymb,amsfonts}
\usepackage{algorithmic}
\usepackage{graphicx}
\usepackage{textcomp}
\usepackage{hyperref} 
\usepackage{xcolor}
\usepackage{caption}
\usepackage{subcaption}
\usepackage{multirow}

\def\BibTeX{{\rm B\kern-.05em{\sc i\kern-.025em b}\kern-.08em
    T\kern-.1667em\lower.7ex\hbox{E}\kern-.125emX}}
\begin{document}

\title{A Security Enhanced Authentication Protocol}

\author{
\IEEEauthorblockN{Sai Sreekar Vankayalapati}
\IEEEauthorblockA{\textit{Computer Science and Engineering} \\
\textit{Indian Institute of Information Technology}\\
Sri City, Chittoor, India \\
saisreekar.v20@iiits.in \\
}

\and
\IEEEauthorblockN{Srijanee Mookherji}
\IEEEauthorblockA{\textit{Computer Science and Engineering} \\
\textit{Indian Institute of Information Technology}\\
Sri City, Chittoor, India \\
srijanee.mookherji@iiits.in \\
}
\and
\IEEEauthorblockN{Vanga Odelu}
\IEEEauthorblockA{\textit{Computer Science and Engineering} \\
\textit{Indian Institute of Information Technology}\\
Sri City, Chittoor, India \\
odelu.vanga@iiits.in \\ 
}

}

\maketitle
\begin{abstract}
Internet of Things (IoT) have gained popularity in recent times. With an increase in the number of IoT devices, security and privacy vulnerabilities are also increasing. For sensitive domains like healthcare and industrial sectors, such vulnerabilities can cause havoc. Thus, authentication is an important aspect for establishing a secure communication between various participants. In this paper, we study the two recent authentication and key exchange protocols. We prove that these protocols are vulnerable to replay attack and modification attack, and also suffer from technical correctness. We then present the possible improvements to overcome the discussed vulnerabilities. The enhancement preserves performance of the original protocols.
\end{abstract}


\maketitle


\section{Introduction} \label{s:intro}
With technology on a rise, Internet of Things (IoT) have become popular. Smart environment with IoT devices are used in almost every field in today's world. It is predicted that by the year 2030, the total number of connected devices would be around 24 billion~\cite{hatton2020}. Smart industries uses IoT devices to collect information about monitoring and tracking of goods, machinery automation and so on. Similarly, healthcare is another domain that has vastly incorporated IoT technologies to efficiently provide medical services to the users. Evidently, these fields contain sensitive and confidential information of the users. Leakage of such data can lead to hazardous consequences. Data Breaches can cause severe operational disruption and financial losses for industries. There have been several incidents of big corporations facing losses due to data leakages, including Tesla where 75000 individuals' private data got leaked~\cite{tesla2023}. Healthcare data leakages are an even bigger issue, because this could be potentially life-threatening for the patients~\cite{challa2018}. Communication of data in these sectors is an extremely sensitive matter that needs to give utmost importance to data privacy. Thus, secure authentication and key exchange plays a major role in providing security to sensitive communications. Recently, Masud $et$ al. proposed authentication protocols for industrial IoT~\cite{masud2021} and healthcare~\cite{masud2022}. However, the works suffer from various security flaws and correctness issues. A secure authentication scheme is important for sensitive domains like smart industry and healthcare. This motivated us to crypt-analyse the works and provide possible modification suggestions in order to address the security flaws and correctness issues. 
\par The paper is further organised as follows:
Table 1~\ref{NotationTable} provides an overview of the mathematical notations used in the paper. Section~\ref{s:protocol1} provides an in-depth analysis of Masud \textit{et} al's.~\cite{masud2021} proposed scheme. In the following Section~\ref{comment1}, crypt-analysis of the protocol is presented. Next, in Section~\ref{protocol2} we explore a similar protocol by Masud \textit{et} al.~\cite{masud2022} and discuss the protocol, followed by it's crypt-analysis in Section~\ref{comment2}. In Section~\ref{s:modif} we present some possible enhancements in order to over come the existing vulnerabilities. Finally, the paper is concluded in Section~\ref{conclusion}.

\section{Review of Masud~\textit{et} al's protocol~\cite{masud2021}} \label{s:protocol1}
In this section, we summarize the recent physically secure and authenticated key establishment protocol proposed by Masud \textit{et} al..~\cite{masud2021}. The scheme aims to establish secret session keys between a User Device (UD) and a Sensor Node (SN) via a Gateway (GW), which it does in three phases, namely the User Registration Phase, the Sensor Node Registration Phase and the Mutual Authentication and Key Agreement Phase. The phases are discussed in detail below. Note that the notations used in this paper are listed in Table \ref{NotationTable}.

\begin{table}[!htb]
\centering
\caption{Notations and Description}\label{NotationTable}
\begin{tabular}{|c|c|} \hline
Notation                  & Description \\ \hline
$P(\cdot)$                       & Physically Unclonable Function \\ \hline
$\oplus, \parallel, h(.)$ & Bitwise XOR, Concatenation, Hash Function respectively \\ \hline
$ID_u, ID_{s_n}$            & Real Identities of User and sensor node respectively \\ \hline
$\beta _u, \beta _{sn}$    & Temporary Identities of user and sensor node respectively \\ \hline
$B_m, H_t, N$             & Biometric, hard token, nonce \\ \hline
$SK$                      & Secret session key \\ \hline
$(C, R)$                  & Challenge response pairs \\ \hline
$K_u, K_{s_n}$              & Random secret keys \\ \hline     
$M_i$ & Message \\ \hline
UD & User device \\ \hline
GW & Gateway \\ \hline
SN & Sensor node \\ \hline
$S_1 , S_2$ & Extraction and helper string for biometrics \\ \hline
\end{tabular}
\end{table}

\subsection{User Registration Phase (Over Secure Channel)} \label{ss:user_reg}
In this phase, the user proves their legitimacy to the GW and registers themselves over a secure channel in the following way:
\begin{enumerate}
    \item User sends their actual identity $(ID_u)$ to the GW as a registration request ($M_1 = ID_u$)
    \item GW generates a challenge $C_u$ that will be used for authentication. In order to address the issues of DoS and desynchronization, it generates additional challenges $C_{uadd} = (C_{u1}, C_{u2}, \ldots, C_{un})$. Finally, the GW sends back $M_2 = C_u, C_{us}$.
    \item UD calculates the responses using its Physically Unclonable Function (PUF) as $R_u = P(C_u)$ and $R_{uadd} = (C_{u1}, C_{u2}, ... C_{un})$ and sends them back to the GW.
    \item The GW generates and sends temporary identity and a random secret key $\beta _u and K_u$ respectively.
    \item The user enters their biometric and attaches a hard token, after which UD computes $X=h(B_m||H_t)$ for future authentication. Lastly, UD stores ${X,\beta _u, K_u}$
    
\end{enumerate}

\subsection{Sensor Node Registration Phase (Over Secure Channel)} \label{ss:sn_reg}

Similar to the User Registration, the sensor nodes also register themselves with the GW over a secure channel through the following steps:

\begin{enumerate}
    \item The GW generates challenges $C_{sn} and C_{snadd} = {C_{sn1}, ... C_{snN}}$ and, along with them, a random secret key $K_sn$ and sends all of them to the sensor node.
    \item SN stores $R_sn$ and computes responses for the challenges $R_{sn}$ and $R_{snadd}$ and sends them to the GW.
    \item The GW generates a temporary ID $\beta _{sn}$ for the sensor node and sends it.
\end{enumerate}

\subsection{Mutual Authentication and Key Agreement Phase (Over Public Channel)} \label{ss:mauth}

The final phase of the protocol is where the GW assigns secret session keys to the user and the sensor node by making use of One Time Pad encryption technique and a hash function in the following way:

\begin{enumerate}
     
    \item UD $\rightarrow$ GW : $M_1 = \langle \beta _u, N_1 \rangle$
    \begin{itemize}
        \item UD verifies the identity of the user using his biometric and a hard token
        \item Generates nonce $N_1$
        \item Finally sends $M_1 = {\beta _u , N_1}$ to the GW.
    \end{itemize}
    
    \item GW $\rightarrow$ UD : $M_2 = \langle N_2, K_u^*, C_u^*, SK_u^*, \beta^{new}_u \rangle$
    \begin{itemize}
    
        \item The GW finds the $\beta _u$ and fetches the respective $K_u$
        \item The GW computes $K_u^* = h(\beta _u || K_u)$ and validates the nonce for freshness.
        \item Now, the GW fetches the challenge response $(C_u, R_u)$ pair corresponding to the user.
        \item The session key is being encrypted using the challenge response pair as $SK_u^* = SK_u \oplus h(C_u||R_u)$. It encrypts the challenges as $C_u^* = C_u \oplus R_u$ 
        \item The GW generates $\beta _u^{new} = h(\beta _u || K_u)$
        \item GW generates $N_2$ and it stores $SK_u, \beta _u^{new}$ and finally the GW sends $M_2 = {N_2, K_u^*, C_u^*, SK_u^*, \beta^{new}_u}$
        
    \end{itemize}
    
    \item UD $\rightarrow$ GW : $M_3 = \langle R^*_u, N_3, \beta _{sn} \rangle$
    
    \begin{itemize}
    
        \item UD validates $N_2$ and it verifies $K_u^*$ by computing $h(\beta _u||K_u)$
        \item UD, then, derives $C_u = C^*_u \oplus K_u$
        \item UD computes $R_u = P(C_u)$ and then it decrypts $SK_u$ as $SK_u = SK_u^* \oplus h(C_u || R_u)$
        \item UD stores $SK_u, \beta^{new}_u$ and calculates $R^*_u = R_u \oplus K_u \oplus SK_u$ and generates a nonce $N_3$
        \item Finally, UD sends back $M_3 = {R^*_u, N_3, \beta _{sn}}$
        
    \end{itemize}

    \item GW $\rightarrow$ SN : $M_4 = \langle N_4, \beta _u, K_{sn}^*, C_{sn}^*, SK_{sn}^*, \beta _{sn}^{new} \rangle$
    
    \begin{itemize}
    
        \item The GW validates $N_u$, and then calculates $R_u \oplus K_u \oplus SK_u$ and verifies it with $R_u^*$ after which the GW computes $K_{sn}^* = h(\beta _{sn} || K_{sn})$.
        \item The GW then fetches $C_{sn}, R_{sn}$ and computes $C_{sn}^* = C_{sn} \oplus K_{sn}$ and $SK_{sn}^* = SK_{sn} \oplus h(C_{sn} || R_{sn)}$.
        \item GW generates a new identity $\beta _{sn}^{new} = h(\beta _{sn} || K_{sn})$
        \item GW sends a $M_4 = {N_4, \beta _u, K_{sn}^*, C_{sn}^*, SK_{sn}^*, \beta _{sn}^{new}}$
        
    \end{itemize}

    \item SN $\rightarrow$ GW : $M_5 = \langle N_5, R^*_{sn} \rangle$
    
    \begin{itemize}
    
        \item SN computes $h(\beta _{sn}||K_{sn})$ and verifies it with $K_{sn}^*$.
        \item Then it validates $N_4$ and if found fresh, it proceeds to derive $C_{sn}$ from $C_{sn}^*$.
        \item SN computes $R_{sn} = P(C_{sn})$. Once it has the responses, it can decrypt $SK_{sn} = SK_{sn}^* \oplus h(C_{sn}||R{sn})$.
        \item SN stores $SK_{sn}, \beta _{sn}^{new}$ and computes $R^*_{sn} = R_{sn} \oplus SK_{sn} \oplus K_{sn}$
        \item Finally, SN sends $M_5 = {N_5, R^*_{sn}}$ to GW
        \item The session ends with the GW validating $N_5$ and $R^*_{sn}$
        
    \end{itemize}    
    
\end{enumerate}

\section{Comment on Masud \textit{et} al.'s Protocol~\cite{masud2021}} \label{comment1}

The proposed scheme performs authentication and delivers encrypted session key through a single random secret key, while using one time pad encryption. However, the protocol suffers from the following vulnerabilities and correctness issues.

\subsection{Hashing a Biometric} \label{prob1}

In the final step of the User Registration Phase \ref{ss:user_reg}, as well as the very first step of the Mutual Authentication Phase \ref{ss:mauth}, biometric $B_m$ is hashed along with $H_t$ and this value is used for verifying the identity of the User. This is not practical as it is infeasible to hash biometrics. The reason for this is that if the user places their, say, fingerprint in a slightly different position the binary value of the biometric will be different, which will generate a new hash value. Thus, an user will not be able to proceed further.

\subsection{User's Session Key is Available as Plain Text} \label{prob2}
\begin{itemize}
    \item When GW sends $M_2 = \langle N_2, K_u^*, C_u^*, SK_u^*, \beta^{new}_u \rangle$, it computes $C_u^*$ as $C_u^* = C_u \oplus R_u$
    \item After receiving $M_2$, UD derives $C_u = C_u^* \oplus K_u$. This is only possible if $(R_u \oplus K_u) = 0$.
    \item In the next step, UD computes $R_u^* = (R_u \oplus K_u) \oplus SK_u$.
    \item If $(R_u \oplus K_u) = 0$ then $R_u^* = SK_u$ and $SK_u$ is available as plain text
\end{itemize}
If it is assumed that $C_u = C_u^* \oplus R_u$ is mistakenly written as $C_u = C_u^* \oplus K_u$, there still prevail some issues, as discussed in the upcoming section. 

\subsection{Updated Temporary Identities Not Being Verified} \label{prob3}
In order to counter Impersonation attacks, the temporary identity $\beta _u$ gets updated and a $\beta ^{new}_u$ and $\beta ^{new}_{sn}$ are generated for the user and the sensor node respectively, after every session. However, after the GW assigns the new identities, the devices store them with no verification of the values. As it is over a public channel, this makes the $\beta ^{new}_u$ and $\beta ^{new}_{sn}$ vulnerable to modification attacks. Let us assume an adversary captures $M_2$ and/or $M_4$ and tries to modify it as follows:

\begin{enumerate}

    \item GW $\rightarrow$ UD : $M_2 = \langle N_2, K_u^*, C_u^*, SK_u^*, \beta^{new}_u \rangle$
    \begin{itemize}
        \item Let's say the Adversary ($\mathcal{A}$) captures $M_2 = \langle \beta^{new}_u, N_2, K_u^*, C_u^*, SK_u^* \rangle $ and modifies it as $M_2^{'} = \langle \beta^{new'}_u, N_2, K_u^*, C_u^*, SK_u^* \rangle$
        \item UD stores the modified value $\beta ^{new'}_u$ with no verification.
    \end{itemize}
    For a future session:
    \begin{itemize}
        \item UD sends their identity $\beta ^{new'}_u$ to GW in $M_1$
        \item There is a mismatch between $\beta ^{new'}_u$ and the stored $\beta ^{new}_u$ which causes the service to disrupt.
        
    \end{itemize}

    \item GW $\rightarrow$ SN : $M_4 = \langle N_4, \beta _u, K_{sn}^*, C_{sn}^*, SK_{sn}^*, \beta _{sn}^{new} \rangle$
    \begin{itemize}
        \item Let's say the Adversary ($\mathcal{A}$) captures $M_4 = \langle N_4, \beta _u, K_{sn}^*, C_{sn}^*, SK_{sn}^*, \beta _{sn}^{new} \rangle$ and modifies it as $M_4^{'} = \langle N_4, \beta _u, K_{sn}^*, C_{sn}^*, SK_{sn}^*, \beta _{sn}^{new'} \rangle$
        \item SN stores the modified value $\beta ^{new'}_{sn}$ with no verification.
    \end{itemize}
    For a future session:
    \begin{itemize}
        \item Assuming that $\beta _{sn}^{new'}$ is public information, UD sends the $\beta _{sn}^{new'}$ that it wants to communicate with to GW in $M_3$
        \item There is a mismatch between $\beta _{sn}^{new'}$ and the stored $\beta _{sn}^{new}$ which leads to service disruption.
    \end{itemize}

\end{enumerate}

\subsection{Nonce Validation Problem Resulting in Replay Attack}
A freshness check is performed on the nonces in the protocol. However, the freshness is validated by checking the randomness of the nonces. As a result, any random number can be used by the Adversary ($\mathcal{A}$) to perform a replay attack. Even if there is a possibility of storing the older nonces, there is high probability that the selected random number is new, thus will be considered as fresh.

\section{Review of Masud \textit{et} al.'s protocol~\cite{masud2022}} \label{protocol2}
A protocol similar to Masud $et$ al.'s~\cite{masud2021} protocol (Section~\ref{s:protocol1}) was proposed by Masud \textit{et} al.~\cite{masud2022} incorporating the following additions:
%



\begin{itemize}
\item This protocol is proposed for the Healthcare industry and has a decentralized blockchain architecture to increase trust between parties. 

\item The information in blockchain is stored as transactions which are further contained in blocks. Each block in the blockchain stores information such as the nonce, parent hash, current hash, timestamp, etc. whereas each transaction comprises of transaction fee, transaction hash, timestamp, input data, nonce, etc. 

\item The protocol has three phases namely, User Registration Phase, Sensor Node Registration Phase and Mutual Authentication and Key Agreement Phase. A smart contract is used in the User Registration and the Sensor Node Registration Phases to establish trust in order to register them onto a blockchain network.

\item The biometric information in the User Registration Phase is handled using a fuzzy extractor, which generates and replicates two strings containing the biometric information in the User Registration Phase and the Mutual Authentication Phase respectively.

\item The additions in the Mutual Authentication and Key Establishment Phase as compared to Masud $et$ al.'s protocol is as follows:
    \begin{itemize}
        \item After the user identity is located and the challenge response pair is fetched from the gateway device's memory, the gateway generates a $J = h(R_u||K_u)$ additionally for the user device to verify the gateway device. 

        \item After the user and gateway authentication, the gateway device generates $G_1 = h(S_1||ID_u)||h(R_u)$ and sends it to the blockchain, along with a transaction hash and contract address, in order to establish a smart contract, after which the blockchain network replies with a success or failure response accordingly.

        \item Similarly, the sensor node also verifies the authenticity of the gateway through $h(K_{sn}||R_{sn})$ and after the gateway and sensor node authentication, the gateway sends $G_2 = h(ID_{sn}||R_{sn})$ to the blockchain network, along with a transaction hash and contract address, in order to establish the smart contract with sensor node.

        \item Finally after all the above steps, the gateway shares one session key with both user and sensor node, unlike the discussed protocol in which two distinct session keys were shared after the authentication. Additionally, new temporary identities are generated and sent to the user and sensor node.
        
    \end{itemize}

\section{Comment on Masud \textit{et} al.'s Protocol~\cite{masud2022}} \label{comment2}

The protocol proposed by Masud $et$ al.~\cite{masud2022} suffers from security flaws similar to that of Masud $et$ al.~\cite{masud2021} (Section~\ref{comment1}). They are discussed as follows: 


\subsection{Updated Temporary Identities Not Being Verified}
At the end of the key establishment, new temporary identities are generated for both the user and sensor node in a similar manner to the first protocol. However, upon receiving them, there is no validation for the new identities being done by UD and SN as they are stored immediately as they are. This makes the new identities susceptible to adversarial modifications that would go unnoticed by the devices and would cause the future sessions to fail.

\subsection{Nonce Validation Problem Resulting in Replay Attack}
Similar to the protocol discussed in Section~\ref{s:protocol1}, freshness of the nonce is validated by checking the randomness of the nonce. Thus, any modified random number would still be considered fresh and would enable the possibility of replay attacks.

\section{Proposed Enhancements} \label{s:modif}

 Masud $et$ al.'s work~\cite{masud2021} suffer from technical correctness error of hashing a biometric along with being vulnerable to replay and modification attack. Whereas, Masud $et$ al.'s work~\cite{masud2022} suffers from vulnerability to replay attack and modification attack. In this section, we discuss the possible modifications for Masud $et$ al.'s~\cite{masud2021,masud2022} works.

\subsection{Handling Biometrics }

In Masud $et$ al.'s work~\cite{masud2021}, biometrics is hashed, which is infeasible. Thus, in order to accommodate the small changes in biometric values and ensure that they can be hashed, a fuzzy extractor \cite{dodis2004} can be used. Two strings $(S_1, S_2)$ can be generated from the biometrics using the fuzzy extractor and those can be used for the further authentication purposes as follows:
\begin{itemize}
    \item In the User Registration Phase:
    \begin{itemize}
        \item At UD: $Gen(B_m) = (S_1, S_2)$ 
        \item $UD \rightarrow GW$: $\delta = h(S_1 || H_t)$. $S_1$ can be concatenated with anything else that is required for authentication. In the case of Masud \textit{et} al's~\cite{masud2021} protocol it is the hard token.
    \end{itemize}
    \item In the Mutual Authentication Phase: 
    Whenever there is a need for biometric authentication, a Replicate function can be used to replicate the $S_1$ as $Rep(B_m, S_2) = S_1$ and this can be hashed and sent again as follows:
        \begin{itemize}
            \item  $S_1 = Rep(B_m, S_2)$
            \item  $\delta = h(S_1 || H_t)$
        \end{itemize}
\end{itemize}

In the enhancement, we use fuzzy extractor to utilize biometrics for login and authentication. This overcomes the technical correctness issue in Masud $et$ al.'s work~\cite{masud2021}.

\subsection{Updating Temporary Identities}

In Masud $et$ al.'s works~\cite{masud2021,masud2022} the temporary identities are updated and sent to the respective parties, which is stored in the device memories without validation. Since the updated identities are sent over insecure channel, an adversary can modify the identities and thus, stop any future communication. In order to overcome the flaws, the following enhancement can be made: 
\par After the end of every session, the gateway computes new temporary identities $\beta _u = h(\beta _u || K_u)$ and $\beta _{sn} = h(\beta _{sn} || K_{sn})$ and they get sent to the respective parties. Before storing the new identities directly, UD and SN can perform a validation to check if the new identities are valid by computing $h(\beta _u||K_u)$ at UD and $h(\beta _{sn}||K_{sn})$ at SN and validating them with $\beta ^{new}_u$ and $\beta^{new}_{sn}$ respectively so that modification attacks are not possible.
In the proposed enhancement, we use cryptographic hash function in order to prevent modification attack.

\subsection{Nonce Validity Check}
In Masud $et$ al.'s works~\cite{masud2021,masud2022}, random nonces are generated along with every message during communication to prevent replay attacks, i.e, if a repeated nonce is found, the communication is disconnected. However, in the discussed protocols, any new random number could pass as being a fresh nonce, which would still make replay attacks possible. Instead of a random nonce, or along with it, the current timestamp could also be sent that would ensure that replay attacks are not possible~\cite{das2018}.


\section{Conclusion} \label{conclusion}
In this paper, we study the authentication and key establishment protocols proposed by Masud $et$ al.~\cite{masud2021,masud2022}. The protocols suffer from technical correctness of hashing biometric and vulnerable to replay attack and modification attack. We present a set of possible enhancement in order to overcome the vulnerabilities. 
\end{itemize}

\bibliographystyle{ieeetr}
\bibliography{SEAP}
\end{document}